\documentclass[epj]{svjour}
\usepackage{graphics}
\usepackage{bm}
\usepackage{here}                    
\usepackage{latexsym}                
\usepackage{amsmath}
\usepackage{amssymb} 

\newcommand{\bra}[1]{\langle #1 | \,}
\newcommand{\ket}[1]{\, | #1 \rangle}

\newcommand{\be}{\begin{equation}}
\newcommand{\ee}{\end{equation}}
\newcommand{\bea}{\begin{eqnarray}}
\newcommand{\eea}{\end{eqnarray}}
\newcommand{\olap}[2]{\langle #1 | #2 \rangle}

\def\unity{\mbox{\rm\bf 1}}

\begin{document}
\title{Provable entanglement and information cost for qubit-based 
quantum key-distribution protocols}
\author{Georgios M. Nikolopoulos, Aeysha Khalique \and Gernot Alber} 
%
\offprints{}          
\institute{Institut f\"ur Angewandte Physik, Technische 
Universit\"at Darmstadt, 64289 Darmstadt, Germany}
\date{Received: date / Revised version: \today}
%
\abstract{Provable entanglement has been shown to be a necessary precondition 
for unconditionally secure key generation in the context of quantum 
cryptographic protocols. We estimate the maximal threshold disturbance up to 
which the two legitimate users can prove the presence of quantum correlations 
in their data, in the context of the four- and six-state quantum 
key-distribution protocols, under the assumption of coherent attacks. 
Moreover, we investigate the conditions under 
which an eavesdropper can saturate these bounds, by means of 
incoherent and two-qubit coherent attacks. 
A direct connection between entanglement distillation and classical advantage 
distillation is also presented.
} 

\authorrunning{G. M. Nikolopoulos {\em et al.}}
\titlerunning{Provable entanglement and information cost for qubit-based  
QKD protocols}

\maketitle
\section{Introduction}
Quantum key-distribution (QKD) protocols exploit qu\-a\-ntum correlations
in order to establish a secret key between two legitimate users
(Alice and Bob). 
In a typical quantum cryptographic scheme, after the transmission
stage Alice and Bob must process their raw key, in order to end up
with identical random keys about which an adversary (Eve) has 
negligible information.
In principle, classical as well as quantum algorithms (distillation
protocols) can be used for this post-processing
\cite{BS,BBCM,M,GL,C,CK,Cascade,DEJ,BDSW,BDSW2}.
In any case, it is necessary for Alice and Bob to estimate the error
rate in their sifted key, for the purpose of detecting the presence
of Eve on the channel.

An important quantity for any QKD protocol
is the threshold disturbance i.e., the maximal disturbance or
{\it quantum bit error rate} (QBER) which can be tolerated by
Alice and Bob for being capable of producing a secret key.
This threshold disturbance quantifies the robustness of the QKD
scheme under consideration against a specific eavesdropping strategy, 
and depends on the algorithm that Alice and Bob are using 
for post-processing their raw key.
Up to date, the robustness of the four-state (BB84) \cite{BB84} 
and the six-state \cite{B}
QKD protocols has been mainly discussed on the basis of the so-called 
Csisz\'ar-K\"orner criterion \cite{CK} and/or incoherent attacks, and various 
bounds have been obtained  \cite{FGNP,BM,CBKG,BKBGC,GW,GW2,AGS,AMG,DBS}. 
Moreover, it is also known that a necessary precondition for  unconditionally 
secure QKD is that the correlations established between Alice and Bob during 
the state distribution cannot be explained in the framework of separable 
states ({\em provable entanglement}) \cite{CLL,AG}.
Clearly, the threshold disturbance up to which this precondition is 
satisfied under the assumption of general coherent (joint) attacks, 
quantifies the ultimate robustness bound of a particular QKD protocol. 

In a recent paper \cite{NA}, we proved that for QKD protocols using two 
mutually unbiased bases, this threshold disturbance for provable entanglement 
(robustness bound) scales with the dimension $d$ of the information carriers 
as $(d-1)/2d$.  
Thus for the BB84 QKD protocol $(d=2)$ \cite{BB84}, Alice and Bob always 
share provable 
entanglement for estimated disturbances below $1/4$.  
Extending our studies, in this paper it is shown that 
the corresponding threshold disturbance for entanglement distillation 
in the context of the six-state QKD protocol \cite{B} is $1/3$. 

Our studies show that even the most powerful eavesdropping attacks are  
not able to disentangle the two legitimate users for estimated disturbances 
below these borders. In other words, Eve is not able to decrease the 
robustness of the protocols. The natural question arises, however, is whether 
and at which cost these disentanglement thresholds can be attained  
in the framework of eavesdropping attacks that maximize Eve's properties 
(information gain and/or probability of success in guessing).  
In this paper we address this open question in the context of incoherent 
as well as two-qubit coherent attacks. 
In particular, we present evidence that in the limit of 
many pairs, coherent attacks might be able to disentangle the two honest 
parties at the lowest threshold disturbance while simultaneously 
maximizing Eve's probability of success in guessing correctly the transmitted 
signal. 

This paper is organized as follows : 
In Section~\ref{basics} we briefly describe the prepare-and-measure 
as well as the associated e\-nta\-ngle\-ment\--\-based versions of the BB84 
and the six-state QKD protocols. The corresponding threshold 
disturbances for provable entanglement (robustness bounds) are 
derived in Section~\ref{n-coherent},  
while in Section~\ref{price} we investigate the cost at which an 
eavesdropper can saturate these bounds. 
A link between entanglement distillation and 
classical advantage distillation protocols is discussed 
in Section~\ref{maurer}. 

\section{Basic facts about BB84 and six-state protocols}
\label{basics}
\noindent
For the sake of completeness, in this section we briefly summarize basic 
facts about the two qubit-based QKD protocols especially in connection 
with their verification-test stage.

\subsection{Prepare-and-measure schemes}
In the prepare-and-measure BB84 protocol \cite{BB84},
Alice sends a sequence of qubits to Bob each of which is randomly
prepared in one of the basis
states $ \{\ket{0},\ket{1}\}$ or $\{\ket{\bar 0},\ket{\bar 1}\}$ 
which are eigenstates of
two maximally conjugated physical variables, namely the two Pauli 
spin operators ${\cal Z}$ and ${\cal X}$.
The eigenstates of
${\cal Z}$, i.e. $\{\ket{0},\ket{1}\}$, and of
${\cal X}$, i.e. $\{\ket{\bar 0},\ket{\bar 1}\}$, 
are related by the Hadamard transformation
\bea
{\cal H}=\frac{1}{\sqrt{2}}\left ( \begin{array}{cc}
\,1\, & \,1\, \\
\,1\, & \,-1\, \\
\end{array}\right ),
\eea
i.e.
$\ket{\bar i}=\sum_j {\cal H}_{ij}\ket{j} ~~(i,j \in\{ 0,1\})$.
In the computational basis $ \{\ket{0},\ket{1}\}$, the Pauli spin operators are
represented by the matrices
\bea
{\cal X}=\left ( \begin{array}{cc}
\,0\, & \,1\, \\
\,1\, & \,0\, \\
\end{array}\right ),\quad
{\cal Y}&=&\left ( \begin{array}{cc}
\,0\, & \,-{\rm i}\, \\
\,{\rm i}\, &  \,0\,\\
\end{array}\right ),\quad
{\cal Z}=\left ( \begin{array}{cc}
\,1\, & \,0\, \\
\,0\, & \,-1\, \\
\end{array}\right ).
\eea
Bob measures the  received  qubits randomly in one of the two bases.
After the transmission stage, Alice and Bob apply a random permutation 
of their data and publicly discuss the bases chosen, 
discarding all the bits where they have selected different bases. 
Subsequently, they randomly select a number of the bits from the remaining
random key (sifted key) and determine their {\em error probability} or QBER.
If, as a result of a noisy quantum channel or of an eavesdropper, the  
estimated QBER is too high the protocol is aborted.
Otherwise, Alice and Bob perform  error correction and privacy 
amplification
with one- or two-way classical communication, in order to obtain a
smaller number of secret and perfectly correlated random bits 
\cite{BS,BBCM,M,GL,C}.

The six-state prepare-and-measure scheme is quite similar to the 
BB84 (four-state) scheme \cite{B}. 
More precisely, Alice and Bob use at random three bases 
namely, the two bases used in the BB84 plus an additional one 
$\{\ket{\bar{\bar 0}},\ket{\bar{\bar 1}}\}$ which corresponds to the  
${\cal Y}$ Pauli operator. In analogy to BB84, the three bases are 
related (up to a global phase) via the transformation 
\bea
{\cal T}=\frac{1}{\sqrt{2}}\left ( \begin{array}{cc}
\,1\, & \,-{\rm i}\, \\
\,1\, & \,{\rm i}\, \\
\end{array}\right ),
\label{Top}
\eea 
i.e.
$\ket{\bar i}=\sum_j {\cal T}_{ij}\ket{j}$ and 
$\ket{\bar{\bar i}}=\sum_j {\cal T}_{ij}^2\ket{j}$ with 
$i,j\in\{0,1\}$.

\subsection{Entanglement-based schemes}
It has been shown that, from the point of view of
an arbitrarily powerful eavesdropper, each one of these two 
prepare-and-measure schemes is equivalent to an e\-nta\-ngle\-ment\--\-based
QKD protocol \cite{BBM,LC,SP,L01,LCA,GP,L01b} .
These latter forms of the protocols offer advantages, in particular
with respect to questions concerning their unconditional security, and
work as follows:
Alice prepares each of, say  $2n$, entangled-qubit pairs in a
particular Bell state \cite{bell}, say
$\ket{\rm\Psi^-}\equiv\frac{1}{\sqrt{2}}(\ket{0_A1_B}-\ket{1_A0_B})$
(where the subscripts $A,B$ refer to Alice and Bob, respectively). This state 
is invariant under any unitary transformation of the form 
${\cal U}_A\otimes{\cal U}_B$.
Alice keeps half of each pair and submits the other half to Bob after 
having applied a random unitary transformation 
chosen either from the set 
$\{\unity, {\cal H}\}$ (two-basis protocol) or from the set 
$\{\unity, {\cal T}, {\cal T}^2\}$ (three-basis protocol).

At the end of the transmission stage, Alice announces publicly the 
transformations she applied on
the transmitted qubits and Bob reverses all of them.
At this stage, in an ideal scenario Alice and Bob would share 
$2n$ pairs in the state $\ket{\rm\Psi^-}^{\otimes 2n}$. 
Due to channel noise and the presence of a possible eavesdropper, 
however, at the end of the transmission stage all the $2n$
entangled-qubit pairs will be corrupted. 
In fact, they will be entangled among themselves as well as
with Eve's probe. 
Thus, the next step for Alice and Bob is to estimate the number of 
singlets among the $2n$ shared pairs (alternatively to estimate the 
fraction of pairs which are in error). To this end, they apply 
a verification test which proceeds as follows: 
Firstly, Alice and Bob permute randomly all the pairs, distributing thus any
influence of the channel noise and the eavesdropper equally among all
the pairs \cite{GL,SP}. 
Afterwards, they randomly select a number (say $n_{\rm c}$) of the pairs 
as check pairs, they measure each one of them {\em separately}  
along a common basis and they publicly compare their outcomes. 
The influence of channel noise or of an eavesdropper is thus 
quantified by the average estimated QBER of the check pairs while,  
assuming that the check pairs constitute a fair sample \cite{fair}, 
the estimated QBER applies also to the remaining, yet
unmeasured, $2n-n_{\rm c}$ pairs.

After the verification test all the check pairs are dismissed and,
if the QBER is too high the protocol is aborted.
Otherwise, Alice and Bob apply an appropriate entanglement 
purification 
protocol (EPP)  with classical one- or 
two-way communication \cite{DEJ,BDSW} on the remaining $2n-n_{\rm c}$ 
pairs, in order to distill a smaller number 
of almost pure entangled-qubit pairs. 
Finally, measuring these  almost perfectly
e\-nta\-ngled\--\-qubit pairs in a common basis, Alice and Bob obtain a 
secret random key, about which an adversary has negligible information.

\subsection{Verification test and confidence level}
\label{ver-sec}
In closing this introductory part of the paper let us recall 
some known basic facts about the verification test which are necessary 
for the subsequent discussion. The reasons for which such a 
classical random sampling procedure 
applies to a quantum scenario have been thoroughly discussed in the 
literature \cite{GL,LC,SP,L01,LCA,GP,L01b}. 
Briefly, the {\em com\-mu\-ting-observables} idea allows 
us to reduce any quantum eavesdropping attack (even a joint one) 
to a classical probabilistic cheating strategy, for which 
classical probability theory can be safely applied 
\cite{LC,LCA}. 
Furthermore, Eve does not know in advance which pairs will be used for 
quality checks and which pairs will contribute to the final key. 
Thus she is not able to treat them differently and the check pairs 
constitute a fair \cite{fair} classical random sample of all the pairs 
\cite{GL,LC,SP}. 
By invoking the verification test therefore the two legitimate users 
can be confident that (with high probability) the estimated error 
rate is also the error rate they would have measured if they 
were able to perform a Bell measurement projecting their pairs 
onto a $2n$-pair Bell basis \cite{LC,LCA,GP}. 
The confidence level is determined by classical random sampling theory 
\cite{RS_book}. 
In particular, the conditional probability that the verification test  
is passed given that Alice and Bob underestimate the error rate in their 
pairs is exponentially small in the sample-size $n_{\rm c}$ 
(i.e, $\sim 2^{-n_{\rm c}}$) \cite{LC,LCA}. 
In other words the probability that Eve 
cheats successfully can be made arbitrarily small by choosing a 
sufficiently large sample.  

\section{Provable entanglement and threshold disturbances}
\label{n-coherent}
\noindent
According to a recent observation, a {\em necessary precondition} for secret 
key distillation is that the correlations established between Alice and Bob
during the state distribution cannot be explained by a separable
state \cite{CLL,AG}. Throughout this work, we consider that Alice and
Bob focus on the sifted key during the post-processing
(i.e., they discard immediately all the polarization data for which
they have used different bases) and that they treat each pair
independently.
Thus, according to the aforementioned precondition, given a particular
value of the estimated QBER (observable), the task of Alice and Bob is to infer
whether they share provable entanglement or not.
Thereby, entanglement is considered to be provable if Alice's and
Bob's correlations cannot be explained by a separable state within the
framework of the protocols (including post-processing) and observables 
under consideration.  

Recently \cite{NA}, for the same post-processing, we estimated the threshold 
disturbance for provable entanglement in the context of two-basis qudit-based 
QKD protocols under the assumption of joint eavesdropping attacks. 
In particular, we showed that for estimated disturbances below $(d-1)/2d$  
(where $d$ is the size of the information carriers), Alice and Bob 
can be confident that they share provable entanglement with probability 
exponentially close to one (see Section~\ref{ver-sec}). 
In this section, for the sake of completeness, we briefly recapitulate 
the main steps of our proof adapted to the BB84 scheme.  
Subsequently, along the same lines, we estimate the corresponding 
threshold disturbance for the six-state QKD scheme. For the sake of 
consistency, we will adopt the e\-nta\-ngle\-ment\--\-based versions of the
protocols. We would like to stress, however, that the estimated 
threshold disturbances characterize both versions of the protocols.  

\subsection{BB84 protocol}
\noindent
Given the unitarity and hermiticity of ${\cal H}$, 
the average disturbance (average error probability per qubit pair), 
that Alice and Bob estimate during the verification test is 
given by \cite{GL,NA,SP}
\bea
D = \frac{1}{2n_{\rm c}}\sum_{b=0,1}
\sum_{j_i; i=1}^{n_{\rm c}}
{\rm Tr}_{A,B}\Big \{
\big [{\cal H}_{AB}^b~
{\cal P}~
{\cal H}_{AB}^b\big ]_{j_i}~
 \rho_{AB}
\Big \},
\label{QBER-pro}
\eea
with the projector \cite{perfect} 
\bea
{\cal P}_{j_i}=\sum_{l=0,1}|l_A,l_B\rangle\langle l_A, l_B|=
\ket{{\rm \Phi}^+}\bra{\rm\Phi^+}+\ket{\rm\Phi^-}\bra{\rm\Phi^-},
\label{QBER-pro2}
\eea
and 
${\cal H}_{AB}^b\equiv{\cal H}_A^b\otimes{\cal H}_B^b$. 
The last equality in (\ref{QBER-pro2}) indicates that the verification 
test is nothing more than a quality-check test of the fidelity of 
the $2n$ pairs with respect to the ideal state $\ket{\rm\Psi^-}^{\otimes 2n}$ 
\cite{GL,LC,SP,L01,LCA,GP,L01b}.
The state  $\rho_{AB}$ in Eq. (\ref{QBER-pro}) denotes
the reduced density operator of Alice and Bob for all $2n$ pairs 
while the index $j_i$ indicates that the 
corresponding
physical observable
refers to the $j_i$-th randomly selected qubit pair.
The powers of the Hadamard transformations ${\cal H}^b$, with
$b\in\{0,1\}$, reflect the fact that
the errors in the sifted key originate from measurements in
both complementary bases which have been selected randomly by Alice
and Bob with equal probabilities.

As we mentioned in Section~\ref{ver-sec} one of the crucial cornerstones 
for the unconditional security of the protocol is that 
Eve does not know in advance which pairs will be used for quality checks 
and which pairs will contribute to the final key. Thus she is not 
able to treat them differently and the check pairs constitute a classical 
random sample of all the pairs \cite{GL,LC,SP,L01}. 
To ensure such a homogenization, Alice and Bob permute all of their 
pairs randomly before the verification stage. 
In view of this homogenization, the eavesdropping attack 
(although a joint one) becomes symmetric on all the pairs \cite{GL,SP} i.e., 
$\rho_{AB}^{(1)} = \rho_{AB}^{(2)} =\cdots = \rho_{AB}^{(2n)}$.
Here, the reduced density operator of Alice's and Bob' s $k$-th
pair is denoted by
$\rho_{AB}^{(k)} = {\rm Tr}_{AB}^{(\not k)}(\rho_{AB})$ and 
${\rm Tr}_{AB}^{(\not k)}$ indicates the tracing (averaging) procedure
over all the qubit pairs except the $k$-th one. 
Accordingly, the average estimated disturbance 
(\ref{QBER-pro}) reads \cite{NA}
\begin{eqnarray}
D = \frac{1}{2}\sum_{b=0}^1 
{\rm Tr}_{A,B}^{(j_1)}
\Big \{
\big [({\cal H}_{A}^b\otimes {\cal H}_{B}^b)~
{\cal P}
~({\cal H}_{A}^b\otimes {\cal H}_{B}^b)
\big]_{j_1}
\rho_{AB}^{(j_1)}
\Big \}\nonumber\\
\label{QBER}
\end{eqnarray}
where ${\rm Tr}_{A,B}^{(j_1)}$ denotes the tracing procedure over the
$j_1$-th qubit pair of Alice and Bob.
So, an arbitrary eavesdropping attack which gives rise to
a particular reduced single-pair state $\rho_{AB}^{(j_1)}$ is 
indistinguishable, from the
point of view of the estimated average disturbance, from
a corresponding collective (individual) attack which results in a decorrelated 
$2n$-pair state of the form $\bigotimes_{j=1}^{2n}\rho_{AB}^{(j)}$.

Our purpose now is to estimate the threshold disturbance $D_{\rm th}$ such 
that for any estimated $D < D_{\rm th}$ Alice and Bob can be confident that 
their correlations cannot have emerged from a separable state. 
To this end let us explore the symmetries underlying the observable 
under consideration i.e., the estimated average QBER.  
According to Eqs.  (\ref{QBER}) and (\ref{QBER-pro2}), 
$D$ is invariant under the transformations 
\begin{eqnarray}
(l,b) &\to& (l\oplus_2 1,b),\nonumber\\
(l,b) &\to& (l,b\oplus_2 1),
\label{Symm1}
\end{eqnarray}
where $\oplus_2$ denotes addition modulo $2$.
This invariance implies that the reduced density 
operators $\rho_{AB}^{(j_1)}$ and
\begin{eqnarray}
\tilde{\rho}_{AB}^{(j_1)} &=&\frac{1}{8}\sum_{g\in{\cal G}_1, h
\in {\cal G}_2} U(h)U(g)\rho_{AB}^{(j_1)}U(g)^{\dagger}U(h)^{\dagger}
\label{rhotilde}
\end{eqnarray}
give rise to the same observed value of the QBER \cite{NA}.
The unitary and hermitian operators appearing in Eq.  
(\ref{rhotilde}) form  
unitary representations of two discrete Abelian groups 
${\cal G}_1 =\{g_1,g_2,g_3,g_4\}$ and ${\cal G}_2 =\{h_1,h_2\}$, and 
are given by 
\begin{eqnarray}
U(g_1) &=& {\cal X}_A\otimes {\cal X}_B,\quad 
U(g_2) = {\cal Z}_A\otimes {\cal Z}_B,\nonumber\\
U(g_3) &=& -{\cal Y}_A\otimes {\cal Y}_B,\quad 
U(g_4) = \unity_A\otimes \unity_B,
\label{Ug}
\end{eqnarray}
and
\begin{eqnarray}
U(h_1) = {\cal H}_A\otimes {\cal H}_B,\quad
U(h_2) = \unity_A\otimes \unity_B.
\end{eqnarray}
Moreover, invariance of the average QBER under 
the symmetry transformations of Eq.  (\ref{Symm1}) induces  
invariance of $\tilde{\rho}_{AB}^{(j_1)}$ under both 
discrete Abelian groups ${\cal G}_1$ and ${\cal G}_2$. 

The key point is now that $\rho_{AB}^{(j_1)}$ and 
$\tilde{\rho}_{AB}^{(j_1)}$ 
differ by local unitary operations and convex summation. Thus 
the density operator $\rho_{AB}^{(j_1)}$ is entangled if 
$\tilde{\rho}_{AB}^{(j_1)}$ is entangled.
Our main problem of determining the values of the QBER for which 
Alice and Bob share provable entanglement can be reduced therefore 
to the estimation of the values of $D$ for which the most general 
two-qubit state $\tilde{\rho}_{AB}^{(j_1)}$
(which is invariant under both Abelian discrete groups) 
is entangled.

The hermitian operators $U(g_1)$ and $U(g_2)$ of the group ${\cal G}_1$
constitute already a {\em complete set of commuting operators}
in the Hilbert space of two qubits and the corresponding eigenstates 
are the Bell states \cite{bell}.
Thus, the most general two-qubit state which is invariant under the
Abelian group ${\cal G}_1$ is given by    
\bea
{\tilde \rho}_{AB}^{(j_1)}&=&\lambda_{00}\ket{\rm\Phi^+}\bra{\rm\Phi^+}+
\lambda_{10}\ket{\rm\Phi^-}\bra{\rm\Phi^-}\nonumber\\
&+&\lambda_{01}\ket{\rm\Psi^+}\bra{\rm\Psi^+}+
\lambda_{11}\ket{\rm\Psi^-}\bra{\rm\Psi^-},
\label{rhoAB-Bell}
\eea
with $\lambda_{\alpha \beta}\geq 0$ and  
\bea
\sum_{\alpha,\beta\in\{0,1\}}\lambda_{\alpha\beta} = 1,
\label{norm}
\eea
while additional invariance under
the discrete group ${\cal G}_2$ implies that
\bea
\quad \lambda_{01}=\lambda_{10}.
\label{const1}
\eea
Thus, the state (\ref{rhoAB-Bell}) with the constraint 
(\ref{const1}) is the most general two-qubit state 
invariant under the Abelian groups ${\cal G}_1$ and 
${\cal G}_2$.

For later convenience let us rewrite the state 
${\tilde \rho}_{AB}^{(j_1)}$ in the computational basis, i.e. 
\bea
{\tilde \rho}_{AB}^{(j_1)}=\frac{1}{2}\left ( \begin{array}{cccc} 
D\,\, & 0\,\, & 0\,\, & G\\
0\,\, & F\,\, & H\,\, & 0\\
0\,\, & H\,\, & F\,\, & 0\\
G\,\, & 0\,\, & 0\,\, & D\\
\end{array}\right ),  
\label{rhoAB}
\eea  
with $F=1-D$ denoting the so-called fidelity, i.e. 
the total probability for Bob to receive the submitted signal 
undisturbed. Furthermore, the remaining parameters are given by 
\bea
D = \lambda_{00} + \lambda_{10},\quad\quad 
F = \lambda_{01} + \lambda_{11},
\nonumber\\
G = \lambda_{00} - \lambda_{10},\quad\quad 
H = \lambda_{01} - \lambda_{11}, 
\label{lambda1-4}
\eea
with $D$ denoting the disturbance (QBER).
In general, the parameters $G$ and $H$ can be expressed in terms of 
the overlaps between different states of Eve's probe and are thus  
intimately connected to the eavesdropping strategy. 
The key point for the subsequent discussion, is that for the 
estimation of 
the threshold disturbance it is not required to know the explicit 
form of 
the ``macroscopic'' parameters $G$ and $H$ and their detailed 
dependences on 
Eve's attack.  
More precisely, using Eqs. (\ref{lambda1-4}), 
the constraints (\ref{norm}) and  (\ref{const1}) read 
\bea
F+D=1
\label{const2a}
\eea 
\bea
F+H=D-G
\label{const2b}
\eea 
respectively, while non-negativity of the eigenvalues 
$\lambda_{\alpha\beta}$ implies 
\bea
D\geq|G|,
\label{const3a}
\eea
\bea
F\geq|H|.
\label{const3b}
\eea

\begin{figure}
\vspace{0.7cm}\hspace{1.0cm}\resizebox{0.75\columnwidth}{!}{%
  \includegraphics{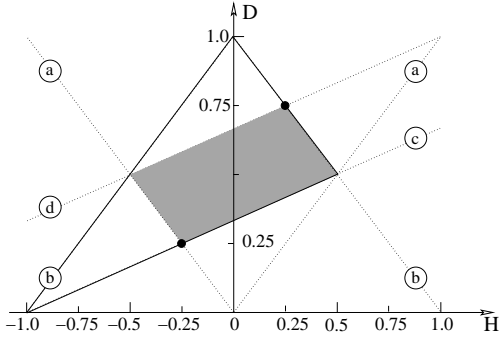}
}
\caption{
BB84 protocol: Region of the independent parameters $D$(QBER) 
and $H$ for which the two-qubit state ${\tilde \rho}_{AB}^{(j_1)}$ is 
separable (shaded region). 
The various constraints that these parameters satisfy 
are indicated by straight dotted lines. Specifically, 
(a) Eq.  (\ref{const4a}); (b) Eq.  (\ref{const3b});  
(c) Eqs. (\ref{const3a}) and (\ref{const2a}), (\ref{const2b}); 
(d) Eqs. (\ref{const4b}) and (\ref{const2a}), (\ref{const2b}). 
The protocol operates in the region which is defined by the solid 
lines.} 
\label{H-G:fig}       
\end{figure}

The possible values of the estimated disturbance for 
which ${\tilde \rho}_{AB}^{(j_1)}$ is entangled 
can be estimated by means of the fully-entangled fraction 
(see \cite{NA}) or the Peres-Horodecki criterion \cite{PH3}.
Using the latter, we have that  ${\tilde \rho}_{AB}^{(j_1)}$ 
is separable {\em if and only if} the inequalities 
\bea
D\geq|H|,
\label{const4a}
\eea
\bea
F\geq|G|,
\label{const4b}
\eea
are satisfied.
As depicted in Fig. \ref{H-G:fig}, these last inequalities combined
with inequalities (\ref{const3a}), (\ref{const3b}) and Eqs.  
(\ref{const2a}), 
(\ref{const2b})  
imply that the symmetrized state ${\tilde \rho}_{AB}^{(j_1)}$ 
is entangled if and anly if the estimated QBER is below $1/4$ 
or above $3/4$.
Given, however, that the states ${\tilde \rho}_{AB}^{(j_1)}$
and $\rho_{AB}^{(j_1)}$ are related via local operations
and convex summation, the original single-pair
state $\rho_{AB}^{(j_1)}$ must also be entangled in the same
regime of parameters. Moreover, the probability that the QBER 
has been underestimated during the verification test is exponentially 
small in $n_{\rm c}$ (see Section~\ref{ver-sec} and related references).
Hence we may conclude that, whenever Alice and Bob detect an average
QBER below $1/4$ (or above $3/4$), they can be confident that
they share entanglement with probability exponentially close to one 
($\sim 1-2^{-n_{\rm c}}$), 
and their correlations cannot have originated from a separable state. 
The necessary precondition for secret-key distillation is therefore 
fulfilled for estimated disturbances within these intervals.

On the contrary, for $1/4\leq D \leq 3/4$ we have that 
${\tilde \rho_{AB}}^{(j_1)}$ is separable. Of course, this does not 
necessarily imply that $\rho_{AB}^{(j_1)}$ is also separable. 
But it does indicate that in this regime of parameters, Alice's and Bob's 
correlations within the framework of the BB84 protocol can be explained 
by a separable state, namely by ${\tilde \rho_{AB}}^{(j_1)}$. 
So, according to \cite{CLL,AG}, this implies that Alice and Bob cannot 
extract a secret key and must abort the protocol. 
From now on we focus on the regime of practical interest $(F\geq D)$, 
where the lowest possible threshold disturbance 
$(D_{\rm th}=1/4)$ is attained for $G=H=-1/4$.

\subsection{Six-state protocol}
\noindent
The threshold disturbances for the six-state protocol can be 
determined in the same way. 
In this case, however, all three bases are used with the 
same probabilities and thus the 
average estimated disturbance (QBER) reads  
\begin{eqnarray}
D =
\frac{1}{3}\sum_{b=0}^2 
{\rm Tr}_{A,B}^{(j_1)}
\Big \{
\big [({\cal T}_{A}^{b}\otimes{\cal T}_B^b)~
{\cal P}~
({\cal T}_{A}^{b\dagger}\otimes{\cal T}_B^ {b\dagger})\big]_{j_1}
\rho_{AB}^{(j_1)}
\Big\}\nonumber\\
\label{QBER6}
\end{eqnarray}
where the unitary (but not hermitian) transformation ${\cal T}$ is 
defined in Eq.  (\ref{Top}).

In analogy to the BB84 protocol, exploiting the symmetries underlying 
Eq.  (\ref{QBER6}) one finds that $D$ is invariant under the 
transformations 
\begin{eqnarray}
  (l,b) &\to& (l\oplus_2 1,b),\nonumber\\
  (l,b) &\to& (l,b\oplus_3 1),\nonumber\\
  (l,b) &\to& (l,b\oplus_3 2),
  \label{Symm2}
\end{eqnarray}
with $\oplus_3$ denoting addition modulo 3. 
Furthermore, the invariance of $D$ under the transformations 
(\ref{Symm2}) 
implies that the reduced density operators $\rho_{AB}^{(j_1)}$ and
\begin{eqnarray}
\tilde{\rho}_{AB}^{(j_1)} &=&\frac{1}{12}\sum_{g\in{\cal G}_1, t
\in {\cal G}_3} 
U(t)U(g)\rho_{AB}^{(j_1)}U(g)^{\dagger}U(t)^{\dagger}
\label{rhotilde6}
\end{eqnarray}
yield the same average QBER.
This latter state is invariant under the discrete A\-be\-li\-an 
groups ${\cal G}_1$ [with elements given in Eq.  (\ref{Ug})] 
and ${\cal G}_3 =\{t_1,t_2,t_3\}$ with elements
\bea
U(t_1) &=& {\cal T}_A\otimes {\cal T}_B,\nonumber\\
U(t_2) &=& {\cal T}_A^2\otimes {\cal T}_B^2,\nonumber\\
U(t_3) &=& \unity_A\otimes \unity_B.
\eea
The most general two-qubit state invariant under the 
Abelian groups ${\cal G}_1$ and ${\cal G}_3$ is now of the 
form (\ref{rhoAB-Bell}), with 
\bea
\lambda_{00}=\lambda_{10}=\lambda_{01}.
\label{const5}
\eea 
Thus, in the computational basis $\tilde{\rho}_{AB}^{(j_1)}$ is 
given by (\ref{rhoAB}) with 
\bea
D &=& 2\lambda_{00},\quad\quad  F=\lambda_{11}+\lambda_{00},\nonumber\\
G &=& 0,\quad\quad\quad \,\, H = \lambda_{00}-\lambda_{11}.
\eea
Accordingly, condition (\ref{const2b}) now reads
\bea
F+H=D,
\label{const6}
\eea
while 
non-negativity of the eigenvalues  $\lambda_{\alpha\beta}$ 
implies inequality (\ref{const3b}) only. 
Finally, applying the Peres-Horodecki criterion one finds that 
$\tilde{\rho}_{AB}^{(j_1)}$ is separable {\em if and only if} 
inequality (\ref{const4a}) is satisfied. 

\begin{figure}
\vspace{0.7cm}\hspace{1.0cm}\resizebox{0.75\columnwidth}{!}{%
  \includegraphics{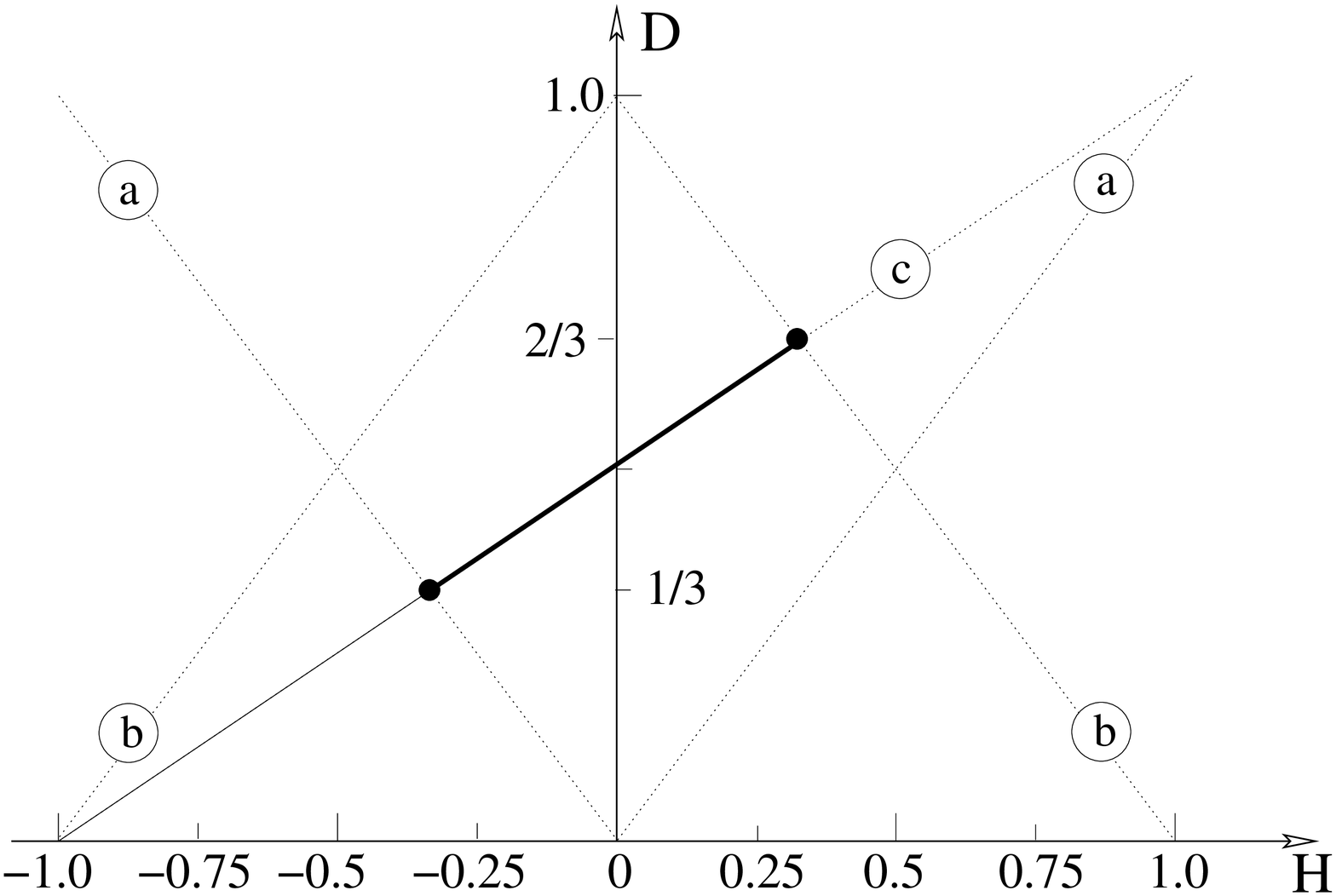}
}
\caption{  Six-state protocol: 
Region of the parameters $D$(QBER) and $H$ for which 
the two-qubit state ${\tilde \rho}_{AB}^{(j_1)}$ is separable 
(thick solid line). 
The various constraints that these parameters satisfy are indicated 
by  straight dotted lines. Specifically, 
(a) Eq.  (\ref{const4a}); (b) Eq.  (\ref{const3b});  
(c) Eqs. (\ref{const2a}) and (\ref{const6}). 
The protocol operates along the solid lines. }
\label{H-G-6:fig}       
\end{figure}

As a consequence of Eqs. (\ref{const2a}), (\ref{const6}) and 
$G=0$,  
there is only one macroscopic independent parameter in our problem, 
say $H$, while combining inequalities (\ref{const3b}) and 
(\ref{const4a}) with 
Eqs. (\ref{const2a}) and (\ref{const6}) we obtain 
that the reduced density operator $\tilde{\rho}_{AB}^{(j_1)}$ 
is separable {\em iff} $1/3\leq D\leq 2/3$ (Fig. \ref{H-G-6:fig}). 
That is, no matter how powerful the eavesdropper is, 
Alice and Bob share always provable entanglement for estimated disturbances 
smaller than $1/3$. The lowest disentanglement border for the 
six-state scheme $(D_{\rm th}=1/3)$ is attained for $H=-1/3$. 
It is also worth noting that, in contrast to BB84, in the six-state 
protocol there is only one disentanglement threshold since for $D>2/3$ 
the protocol is not valid.

As expected, the bound for the six-state protocol is higher than 
the one for the BB84 protocol. In fact, as a consequence of the  high 
symmetry of the six-state protocol, the disentanglement area of the 
BB84 scheme (shaded region in Fig. \ref{H-G:fig}) shrinks to a line in 
Fig. \ref{H-G-6:fig} (thick line). As will be seen later on, this 
``degeneracy'' affects significantly the options of a potential 
eavesdropper in the framework of the six-state protocol, increasing 
thus the robustness of the protocol.
 
\section{The price of disentanglement} 
\noindent
\label{price} 
In QKD issues, Eve's attack is usually optimized by maximizing her 
Shannon information (or the probability of her guessing correctly 
Alice's bit-string) conditioned on a fixed disturbance. 
Given, however, that the unconditional security of the BB84 and six-state 
cryptographic schemes is beyond doubt, Eve might be willing to 
reduce the robustness 
of the protocols to the lowest possible level while simultaneously 
maximizing any of her properties \cite{AGS}. 
Thus, what remains to be clarified now is the cost at which Eve can 
saturate the lowest disentanglement threshold $D_{\rm th}$, in terms 
of her 
information gain and probability of correct guessing.
To this end, we have to consider in detail the eavesdropping attack 
on the BB84 and the six-state protocols.
  
Such an investigation, however, is practically feasible only in the 
context of attacks on a few qubits. 
As the number of attacked qubit-pairs increases the complete treatment 
of the problem becomes intractable due to the large number of independent 
parameters involved. In this section we will focus on incoherent and 
two-qubit coherent attacks. The disentanglement of Alice and Bob in the 
framework of incoherent attacks has been extensively 
studied in the literature \cite{GW,GW2,AGS,AMG,DBS}. 
In most of these studies, however, Eve's attack 
is by default optimized to provide her with the maximal Shannon 
information. On the contrary, here we give Eve all the flexibility to 
adjust her parameters in order to break entanglement between Alice and Bob 
and simultaneously maximize her properties. Finally, for the two QKD 
protocols under consideration, we are not aware of any related previous 
work on disentanglement in the context of coherent attacks.
 
\subsection{BB84 protocol}
\noindent
\subsubsection{Incoherent attacks}
\label{incoherent}
\begin{figure}
\vspace{0.7cm}\hspace{1.0cm}\resizebox{0.75\columnwidth}{!}{%
  \includegraphics{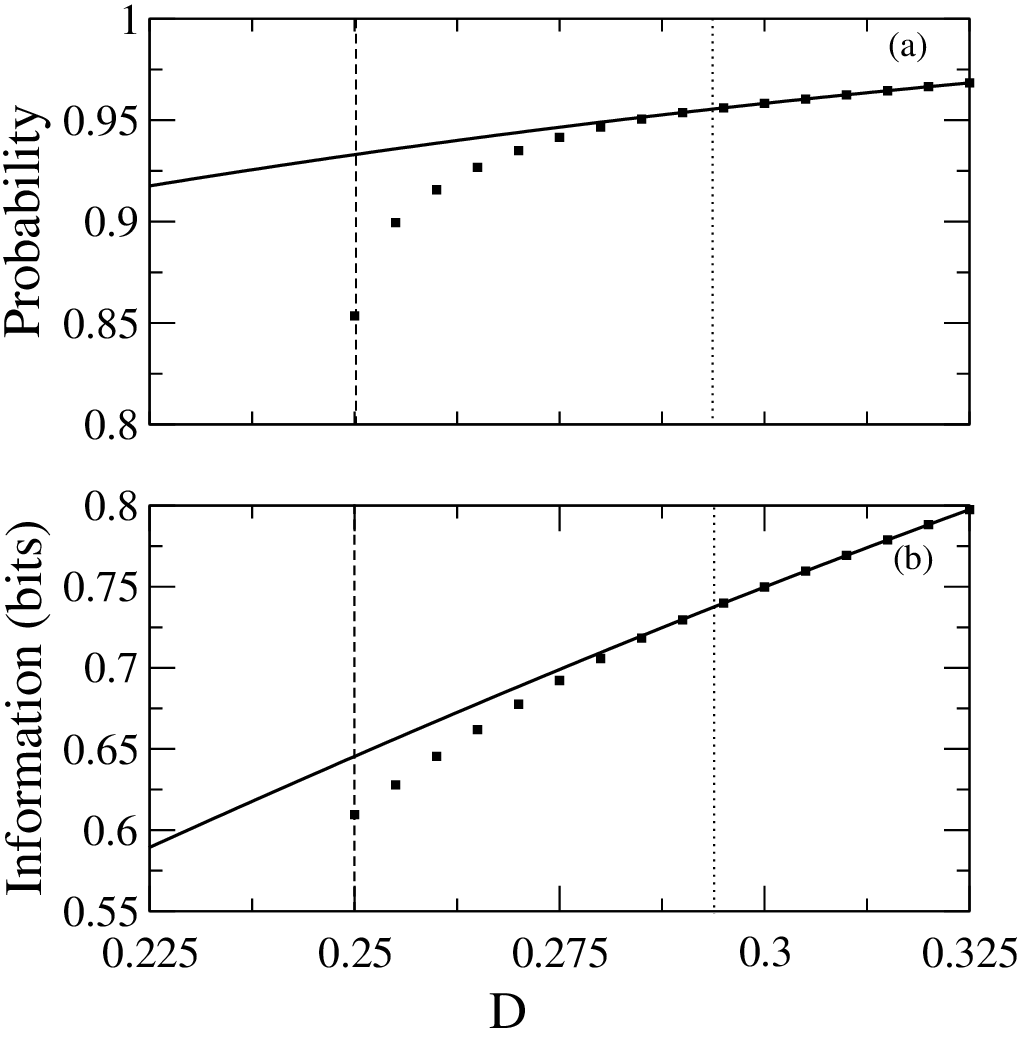}
}
\caption{BB84 protocol --- Incoherent attacks : 
(a) Eve's probability of guessing correctly the transmitted 
message as a function of disturbance $D$. 
The solid line corresponds to an attack that maximizes Eve's 
probability of success in guessing, while each square 
denotes the corresponding probability for an attack which in addition,  
disentangles Alice and Bob at the specific disturbance. 
(b) As in (a) but for Eve's information gain. 
The vertical dotted lines correspond to the solid curves, and denote 
the 
disturbance $D^{(1)}\approx30\%$ up to which Alice and Bob share an entangled 
state. The vertical dashed lines denote the lowest disentanglement 
threshold 
disturbance $D_{\rm th}=1/4$ which can be attained in the context of 
general coherent 
attacks and intercept-resend strategies.}
\label{incoherent:fig}       
\end{figure}

\noindent
Incoherent attacks belong to the class of the so-called 
single-qubit or individual attacks, 
where Eve manipulates each transmitted qubit individually. 
To this end, she attaches a single probe 
(initially prepared in e.g. state $\ket{0_E}$) to each transmitted 
qubit and 
lets the combined system undergo a unitary transformation of the form 
\cite{FGNP,RMP,CG}
\bea
\ket{0_B}\otimes\ket{0_E}\rightarrow\sqrt{F}\ket{0_B}\otimes
\ket{\phi_0}+\sqrt{D}\ket{1_B}\otimes\ket{\theta_0},\nonumber\\
\ket{1_B}\otimes\ket{0_E}\rightarrow\sqrt{F}\ket{1_B}\otimes
\ket{\phi_1}+\sqrt{D}\ket{0_B}\otimes\ket{\theta_1},
\label{IncohUnitary}
\eea
with $F$ and $D$ being the fidelity and disturbance respectively, 
while $\ket{\phi_j}$ and $\ket{\theta_j}$ 
are normalized states of Eve's probe when Bob receives the transmitted 
qubit 
undisturbed (probability $F$) and disturbed (probability $D$), 
respectively. 
Applying unitarity and symmetry conditions on this transformation one 
finds 
that the states $\ket{\phi_j}$ are orthogonal to the states 
$\ket{\theta_j}\quad (j\in\{0,1\})$, while the 
overlaps $\olap{\phi_0}{\phi_1}$ and $\olap{\theta_0}{\theta_1}$ are 
real-valued  \cite{FGNP,RMP,CG}. 
Thus, an incoherent attack can be described by the four parameters 
satisfying Eqs. (\ref{const2a}), (\ref{const2b})  
(\ref{const3a}) and (\ref{const3b}) with $H=-F\olap{\phi_0}{\phi_1}$ 
and 
$G=-D\olap{\theta_0}{\theta_1}$. In other words, there are only two 
independent parameters and by fixing one of them, 
say $D$, one is able to determine any property of the attack. 
In Figs. \ref{incoherent:fig}, we present Eve's 
optimal information gain and probability of success in guessing the 
transmitted qubit correctly as functions of the disturbance 
(solid line). 
The optimization is performed in the usual way, i.e. for a 
fixed disturbance $D$, Eve's mutual information with Alice is 
maximized 
\cite{FGNP,CG}. 
It is also known that such an optimized strategy disentangles the 
qubits of 
Alice and Bob at $D^{(1)}\approx 30\%$ (vertical dotted line)\cite{GW},
which is well above $D_{\rm th}= 25\%$. Thus, the natural question 
arises is whether, under the assumption of incoherent attacks, Eve can 
saturate 
the lowest possible disentanglement border $D_{\rm th}$ and if yes, at 
which cost of information loss.

To answer this question, for a fixed disturbance $D$, 
we calculated numerically all the possible 
values of $G$ and $H$ which are consistent with the constraints 
(\ref{const2a})-(\ref{const3b}) and which yield 
a separable state of Alice and Bob. 
In general, at any given disturbance there is more 
than one combination of values of $G$ and $H$ which fulfill all these 
constraints. 
For each of these combinations, we calculated  
Eve's information gain and her probability of correct guessing 
\cite{FGNP,CG}. 
The results presented as squares in Figs. \ref{incoherent:fig}, 
refer 
to those combinations of parameters which, not only disentangle 
the two honest parties for a particular disturbance $D$, but which   
simultaneously maximize Eve's property as well.
Clearly, for disturbances close to $D_{\rm th}$, the two strategies are 
not 
equivalent since they yield substantially different results.   
In other words, an optimal incoherent attack that maximizes Eve's 
information gain is certainly not the one which achieves the 
lowest possible robustness bound. 
Furthermore, our simulations show that saturation of $D_{\rm th}=1/4$ is 
feasible at the cost of $\sim 4\%$ less information gain of Eve or 
equivalently at the cost of $\sim 7.44\%$ less probability of success 
in guessing. 

\subsubsection{Two-qubit coherent attacks}
\label{2-coherent}
\noindent
In a two-qubit coherent attack, Eve attaches one probe to two of the 
qubits sent by 
Alice. Let $\ket{m_B}$ with $m\in\{0,1,2,3\}$, be the message sent 
from 
Alice to Bob in binary notation. 
The combined system then undergoes a unitary transformation of the 
form \cite{CG}
\bea 
\left (\begin{array}{c}
\ket{0_B} \\
\ket{1_B} \\
\ket{2_B} \\
\ket{3_B}
\end{array}\right )\otimes\ket{0_E}\rightarrow 
{\cal E}\otimes
\left (\begin{array}{c}
\ket{0_B} \\
\ket{1_B} \\
\ket{2_B} \\
\ket{3_B} 
\end{array}\right ),
\label{CohUnitary}
\eea
where ${\cal E}$ is a $4\times 4$ matrix which contains 
normalized states in the Hilbert space of Eve's probe  
\bea
{\cal E}\equiv 
\left (\begin{array}{cccc}
\sqrt{\alpha}\ket{\phi_0} & \sqrt{\beta}\ket{\theta_0} & 
\sqrt{\beta}\ket{\omega_0} & \sqrt{\gamma}\ket{\chi_0} \\
\sqrt{\beta}\ket{\theta_1} & \sqrt{\alpha}\ket{\phi_1} & 
\sqrt{\gamma}\ket{\chi_1} & \sqrt{\beta}\ket{\omega_1} \\
\sqrt{\beta}\ket{\omega_2} & \sqrt{\gamma}\ket{\chi_2} & 
\sqrt{\alpha}\ket{\phi_2} & \sqrt{\beta}\ket{\theta_2} \\
\sqrt{\gamma}\ket{\chi_3} & \sqrt{\beta}\ket{\omega_3} & 
\sqrt{\beta}\ket{\theta_3} & \sqrt{\alpha}\ket{\phi_3} 
\end{array}\right ). \nonumber
\eea 
The states $\phi_j$, $\theta_j$, $\omega_j$ and $\chi_j$ denote  
Eve's probe states in cases in which Bob receives all the transmitted 
qubits 
undisturbed, one qubit disturbed or both transmitted qubits 
disturbed. 

Applying unitarity and symmetry conditions on Eq.  
(\ref{CohUnitary}), 
the problem can be formulated in terms of 
the following four mutually orthogonal subspaces \cite{CG}
\bea
S_\phi &=& \{\phi_0,\,\phi_1,\,\phi_2,\,\phi_3\},\quad\quad
S_\chi =  \{\chi_0,\,\chi_1,\,\chi_2,\,\chi_3\},\nonumber\\
S_\theta &=&  \{\theta_0,\,\theta_1,\,\theta_2,\,\theta_3\},
\quad\quad\,\,\,
S_\omega = \{\omega_0,\,\omega_1,\,\omega_2,\,\omega_3\},\nonumber
\eea
while all the overlaps between the various states within each of these 
subspaces 
are real-valued. 
Thus, Eve is able to infer with certainty whether Bob has received 
both 
qubits undisturbed $(S_\phi)$, one qubit disturbed 
$(S_{\theta,\omega})$ 
or both qubits disturbed $(S_{\chi})$. These events occur with 
probabilities 
$\alpha$, $2\beta$ and $\gamma$, respectively.
It can be shown that a general coherent two-qubit attack can be 
described in 
terms of five independent parameters \cite{CG}. 
The average reduced density matrix for Alice and Bob is 
then of the form (\ref{rhoAB}), with $F=\alpha+\beta$, 
$D=\beta+\gamma$, 
$H=-(\alpha\olap{\phi_0}{\phi_1}+\beta\olap{\theta_0}{\theta_2})$, 
$G=-(\gamma\olap{\chi_0}{\chi_1}+\beta\olap{\theta_0}{\theta_1})$,  
satisfying the constraints (\ref{const2a}), 
(\ref{const2b}),  (\ref{const3a}) and (\ref{const3b}). 

\begin{figure}
\vspace{0.7cm}\hspace{1.0cm}\resizebox{0.75\columnwidth}{!}{%
  \includegraphics{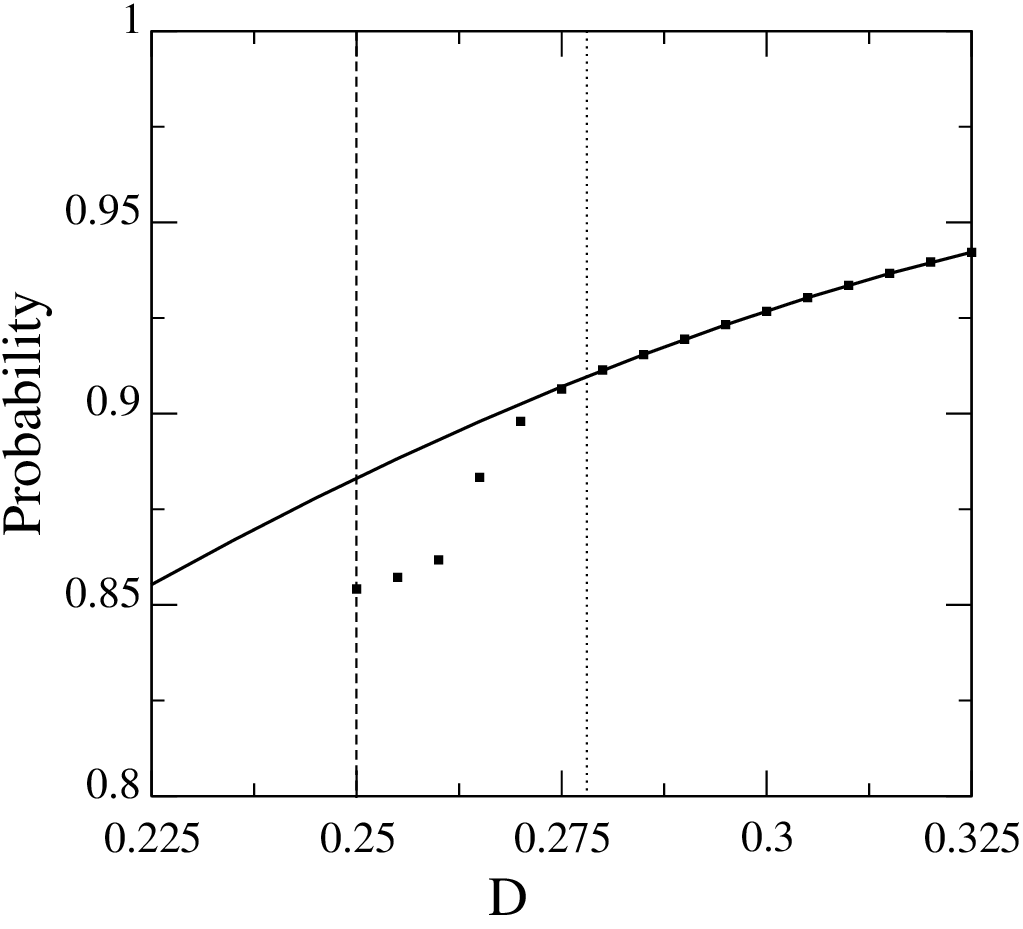}
}
\caption{BB84 protocol --- Two-qubit coherent attacks : 
Eve's probability of guessing correctly a two-bit transmitted 
message as a function of disturbance $D$. 
The solid line corresponds to an attack that maximizes Eve's 
probability of success in guessing only, while each square 
denotes the corresponding probability for an attack that, in 
addition,  
disentangles Alice and Bob at the specified disturbance. 
The vertical dotted line corresponds to the solid curve, and denotes 
the 
disturbance $D^{(2)}\approx28\%$ up to which Alice and Bob share an entangled 
state. 
The vertical dashed line denotes the lowest possible disentanglement 
threshold disturbance 
$D_{\rm th}=1/4$ that can be attained in the context of general coherent 
attacks and 
intercept-resend strategies.}    
\label{coherent:fig}     
\end{figure}

Compared to an incoherent attack, 
a two-qubit coherent attack can improve the probability that Eve 
guesses correctly the whole two-bit message sent by Alice to 
Bob \cite{CG}. 
Eve's optimal probability of success in guessing is plotted in 
Fig. \ref{coherent:fig} (solid line), as a function of disturbance $D$. 
This curve has been obtained by maximizing Eve's probability of 
success in guessing conditioned on a fixed disturbance $D$. 
For such an optimal attack, we found numerically that Alice and 
Bob share entanglement up to disturbances of the order of 
$D^{(2)}\approx 28\%$ (dotted vertical line). 
This is in contrast to the bound $D^{(1)}\approx 30\%$ attained in 
an optimal incoherent attack.  Furthermore, we also found that Eve is 
able to saturate the lowest possible robustness bound 
(dashed vertical line), at the cost of $\sim 3\%$ less 
probability of success in guessing. 
This loss of Eve's probability in guessing is substantially smaller 
than the corresponding loss for incoherent attacks $(\sim 7.44\%)$. 
Thus, it could be argued that a two-qubit coherent attack which is 
optimized 
with respect to the probability of guessing only, is very close to an 
optimal coherent attack which also disentangles Alice and Bob at 
$D_{\rm th}=1/4$.    
The reason is basically that in a two-qubit coherent attack each one 
of 
the two independent macroscopic parameters $G$ and $H$ can be 
expressed in 
terms of two different overlaps whereas in incoherent attacks the 
corresponding dependences involve a single overlap only. 
In a coherent attack Eve has therefore more 
possibilities enabling her to push the disentanglement border 
towards the lowest possible value, while simultaneously maximizing 
her probability of guessing correctly the transmitted message.

\subsection{Six-state protocol} 
\noindent
So far, we have considered incoherent and coherent attacks in the 
context of the BB84 protocol where Eve's attack is determined by a 
set of two macroscopic parameters $(G,H)$. 
These two independent parameters give a considerable flexibility 
to Eve since at a given disturbance there exists 
a variety of physically allowed attacks. 
This fact is also reflected in Fig. \ref{H-G:fig} where, 
for a specific disturbance, Alice and Bob can be disentangled 
for different values of $H$ (and therefore of $G$).

In the highly symmetric six-state protocol, however, the situation 
is much simpler. In fact, the high symmetry of 
the protocol reduces significantly the options of an eavesdropper 
since 
there is only one independent  macroscopic parameter in our problem, 
namely 
$H$. Moreover, the analysis of the attacks under consideration becomes 
rather straightforward \cite{PG}. In particular, for incoherent 
attacks  
$G=-D\olap{\theta_0}{\theta_1}=0$ which indicates that Eve has 
full information about the disturbed qubits received by Bob. 
However, as depicted in Fig. \ref{H-G-6:fig}, at a given value of $D$  
there is a unique value of $H$ consistent 
with the laws of quantum mechanics. 
It is determined by Eqs. (\ref{const2a}) and (\ref{const6}) 
[line (c) in Fig. \ref{H-G-6:fig}]. Similarly, for the two qubit 
coherent attack we have  $\olap{\chi_0}{\chi_1}=\olap{\theta_0}{\theta_1}=0$ 
and thus $G=0$, whereas 
$H=-(\alpha\olap{\phi_0}{\phi_1}+\beta\olap{\theta_0}{\theta_2})=
-(\alpha-\gamma)=2D-1$. 
As a result, for both incoherent and two-qubit coherent attacks, the 
physically allowed 
attack is the one that maximizes Eve's probability of guessing and 
simultaneously 
disentangles Alice and Bob at a given disturbance. 
It is sufficient for Eve therefore to optimize her attack with respect 
to 
her probability of correct guessing in order to disentangle Alice and 
Bob at the lowest 
possible disturbance.

\section{Entanglement and intrinsic information}
\label{maurer}
\noindent
So far, we have discussed for both the four- and six-state 
protocols the maximal disturbance up to which Alice and Bob share 
entanglement. Clearly, this bound indicates that in principle 
secret-key generation is feasible by means of a quantum purification 
protocol. 
In this section we show that, at least in the context of incoherent 
attacks, 
a two-way classical protocol, the so-called advantage distillation 
protocol, exists which can tolerate precisely the same amount of 
disturbance 
as a quantum purification protocol.    
 
To this end, we adopt Maurer's model for classical key agreement 
by public discussion from common information 
\cite{M}. Briefly, in this classical scenario, 
Alice, Bob and Eve, have access to {\em independent} realizations of 
random variables $X, Y$ and $Z$, respectively, jointly distributed 
according to $P_{XYZ}$. 
Furthermore, the two honest parties are connected by a
noiseless and authentic (but otherwise insecure) channel. 
In the context of this model, Maurer and Wolf have shown that a useful 
upper bound for the secret-key rate $S(X;Y||Z)$ 
is the so called intrinsic information $I(X;Y\downarrow Z)$ which is 
defined as 
\bea
I(X;Y\downarrow Z) = \min_{Z \rightarrow \bar Z}\{I(X:Y|Z)\},\nonumber 
\eea
where $I(X:Y|Z)$ is the mutual information between the variables 
$X$ and $Y$ conditioned on Eve's variable $Z$, while the minimization 
runs over all the possible maps $Z\to \bar{Z}$ \cite{MW}. 

For our purposes, we can link this classical scenario to a quantum one.
More precisely, the joint distribution $P_{XYZ}$ can be thought of as 
arising 
from measurements performed on a quantum state 
$\ket{\rm\Psi_{ABE}}$ shared between Alice, Bob and Eve.  
We have, however, to focus on incoherent attacks where 
Eve interacts individually with each qubit and performs any measurements 
before reconciliation. Thus, at the end of such an attack  
the three parties share independent 
realizations of the random variables $X$, $Y$ and $Z$. Accordingly,   
the resulting mixed state after tracing out Eve's degrees 
of freedom  is of the form (\ref{rhoAB}) where 
$H=-F\olap{\phi_0}{\phi_1}$ and $G=-D\olap{\theta_0}{\theta_1}$. 
It turns out \cite{GW2} that the random variables $X$ and $Y$ are 
symmetric 
bits whose probability of being different is given by 
${\rm Prob}[X\neq Y]=D$ whereas 
Eve's random variable consists of two bits $Z_1$ and $Z_2$. 
The first bit $Z_1=X\oplus_2 Z$ shows whether Bob has 
received the transmitted qubit disturbed $(Z_1=1)$ or undisturbed 
$(Z_1=0)$.   
The probability that the second bit $Z_2$ indicates correctly the 
value of 
the bit $Y$ is given by 
\bea
{\rm Prob}[Z_2=Y]=\delta=
\frac{1+\sqrt{1-\olap{\phi_0}{\phi_1}^2}}{2}.
\label{delta}
\eea

As has been shown by Gisin and Wolf \cite{GW2}, 
for the scenario under consideration secret key agreement is always 
possible 
{\em iff} the following condition holds
\bea
\frac{D}{1-D}< 2\sqrt{(1-\delta)\delta}.
\label{delta2}
\eea 
More precisely, one can show that if the above condition is not 
satisfied, 
the intrinsic information vanishes whereas, in any other case 
there exists a classical protocol that can provide Alice and Bob 
with identical keys about which Eve has negligible information. 
Such a protocol, for instance is the so-called 
advantage distillation protocol which is described in 
detail elsewhere \cite{M}.  

In our case now, considering that Eve has adjusted the parameters in 
her attack to disentangle Alice and Bob at the lowest possible 
disturbance, 
Eq.  (\ref{delta}) yields for the two protocols   
\bea
\delta=\left\{ 
\begin{array}{ll}
\frac{3+2\sqrt{2}}{6}\quad & \textrm{BB84 protocol}\\
\frac{2+\sqrt{3}}{4}\quad & \textrm{six-state protocol}.
\end{array} \right.
\nonumber
\eea
Using these values of $\delta$ in Eq.  (\ref{delta2}) one then 
obtains 
bounds that are precisely the same with the threshold disturbances 
for provable entanglement we derived in Section \ref{n-coherent}. 
In other words we have shown that, as long as Alice and Bob are 
entangled, a classical advantage distillation protocol is capable 
of providing them with a secret key, provided Eve restricts herself 
to individual attacks only (see also \cite{AMG,DBS} for similar 
results). 

This result is a manifestation of the link between quantum 
and secret correlations in both four- and six-state QKD 
protocols \cite{CLL,AG}. 
For the time being, the validity of this equivalence between classical 
and quantum distillation protocols is restricted to 
individual attacks only. 
Investigations of tomographic QKD protocols have 
shown, however, that such an equivalence is invalid for coherent attacks 
\cite{Kas03}.  

\section{Concluding remarks}
\label{conclusions}
\noindent
We have discussed provable entanglement in the framework of the BB84 and the 
six-state QKD protocols under the assumption of coherent(joint) attacks. 
In particular, we have shown that the threshold disturbances for provable 
entanglement are $1/4$ and $1/3$ for the four- and six-state QKD 
protocols, respectively. 
Perhaps surprisingly, these borders coincide with the disentanglement borders 
associated with the standard intercept-resend strategy \cite{IR,IR2}. 
Here we have shown, however, that even the most powerful eavesdropping 
attacks (which are only limited by the fundamental 
laws of quantum theory), are not able to push these disentanglement borders 
to lower disturbances. In other words, for the two protocols under 
consideration, any eavesdropping attack which disentangles Alice and Bob 
gives rise to QBERs above $1/4$ (BB84) and $1/3$ (six-state).  
Hence, for estimated disturbances below these borders the two honest 
parties can be confident (with probability exponentially close to one) 
that their quantum correlations cannot be described 
in the context of separable states and can be explored 
therefore for the extraction of a secret key.

In particular, for the entanglement-based versions of the 
protocols such a secure key can be obtained after
applying an EPP which purifies the qubit pairs shared between Alice
and Bob.
Nevertheless, for the prepare-and-measure forms of the protocols the 
situation is more involved.
To the best of our knowledge, the highest tolerable error rates that 
have 
been reported so far in the context of the prepare-and-measure 
BB84 and six-state schemes are close to $20\%$ and $27\%$, 
respectively \cite{GL,C}.
These best records are well below the corresponding threshold 
disturbances 
we obtained in this work.  
Thus, an interesting open problem is the development of 
prepare-and-measure protocols which bridge the remaining 
gap and are capable of generating
a provably secure key up to $25\%$ and $33.3\%$ bit error rates.
In view of the fundamental role of entanglement in secret key
distribution such a development appears to be plausible. 
For this purpose, however, construction of new appropriate
EPPs with two-way classical communication, which are consistent with the
prepare-and-measure schemes, is of vital importance.

Furthermore, we have investigated the cost of information loss at which an 
eavesdropper can saturate these bounds in the context of symmetric 
incoherent and two-qubit coherent attacks.  
We have found that for the highly symmetric six-state scheme, there 
is always a unique eavesdropping attack which disentangles Alice and Bob 
at a fixed disturbance (above $1/3$) and simultaneously 
maximizes Eve's information gain and/or probability of guessing.  
For the BB84 protocol, however, the situation is 
substantially different. Specifically, an attack which maximizes any 
of Eve's properties (information gain or probability of success in 
guessing) is not necessarily also the one that yields the lowest 
possible robustness bound.
In fact, if Eve aims at reducing the robustness of the BB84 protocol 
she
has to accept less information gain and probability of correct 
guessing.
Nevertheless, our simulations show that for a two-qubit coherent 
attack 
this cost is substantially smaller than the cost for an incoherent 
attack.
We conjecture therefore that, for coherent attacks on a larger number 
of
qubits, the strategy that maximizes Eve's probability of success in 
guessing,
is also the one that defines the lowest possible disentanglement 
threshold.

In closing, it should be stressed that the bounds we have 
obtained throughout this work depend on the post-processing that 
Alice and Bob apply. 
In particular, they rely on the complete omission of 
any polarization data from the raw key that involve different bases 
for Alice and Bob as well as on the individual manipulation 
of each pair of (qu)bits during the post-processing. In other 
words only one observable is estimated, namely the disturbance or 
QBER. 
If some of these conditions are changed, also the threshold
disturbances may change. In this context it was demonstrated recently 
that with the help of entanglement witnesses which are constructed from 
the data of the raw key, the detection of
quantum correlations between Alice and Bob is feasible even for QBERs
above the bounds we have obtained here \cite{CLL}.

\section{Acknowledgments}
\noindent
Stimulating discussions with Nicolas Gisin and Norbert L\"utkenhaus 
are gratefully acknowledged. This work is supported by
the EU within the IP SECOQC.



\begin{thebibliography}{000}
\bibitem{fair}
In general, a logarithmic scaling of the size of the 
random sample with the length of Alice's and Bob's key, 
seems to be sufficient for security issues. 
See Ref. \cite{LCA} for a rigorous proof
\bibitem{perfect}
Note that in the absence of noise and eavesdropping 
each pair of qubits shared between Alice and Bob is 
in the Bell state $\ket{\rm\Psi^-}$ \cite{bell}. 
Thus, in this ideal scenario, Alice and Bob obtain 
perfectly anticorrelated measurement results 
whenever they perform their measurements along 
the same basis
\end{thebibliography}
\end{document}